\documentclass[aps,reprint,onecolumn, showpacs,notitlepage,nofootinbib, natbib, superscriptaddress]{revtex4-1}
\usepackage{graphicx}
\usepackage{epstopdf, epsfig}
\usepackage{amsbsy}

\usepackage[normalem]{ulem}
\usepackage{cancel}
\usepackage[utf8]{inputenc}
\AtBeginDocument{\renewcommand{\omega}{V}}

\usepackage{xcolor}
\usepackage{float}

\usepackage[english]{babel}
\usepackage{lipsum}
\usepackage{amssymb,amsfonts,amsmath}

\makeatletter
\newcommand\footnoteref[1]{\protected@xdef\@thefnmark{\ref{#1}}\@footnotemark}
\makeatother

\makeatletter
\newcommand{\customlabel}[2]{%
\protected@write \@auxout {}{\string \newlabel {#1}{{#2}{}}}}
\makeatother

\begin{document}
\title{A thin-film equation for a viscoelastic fluid, and its application to the Landau-Levich problem}

\author{Charu Datt}
\affiliation{Physics of Fluids Group, Faculty of Science and Technology, University of Twente, P.O. Box 217, 7500 AE Enschede, The Netherlands
}

\author{Minkush Kansal}
\affiliation{Physics of Fluids Group, Faculty of Science and Technology, University of Twente, P.O. Box 217, 7500 AE Enschede, The Netherlands
}

\author{Jacco H. Snoeijer} 
\affiliation{Physics of Fluids Group, Faculty of Science and Technology, University of Twente, P.O. Box 217, 7500 AE Enschede, The Netherlands
}

\begin{abstract}
Thin-film flows of viscoelastic fluids are encountered in various industrial and biological settings. The understanding of thin viscous film flows  in Newtonian fluids is {very} {well developed}, {which for a large part is due to the so-called thin-film equation.} This equation, a single partial differential equation describing the height of the film, is a significant simplification of the Stokes {equation effected} by the lubrication approximation which exploits the thinness of the film. There is no such established equation for viscoelastic fluid flows. Here we derive the thin-film equation for a second-order fluid, and use it to study the classical Landau-Levich dip-coating problem. We show how viscoelasticity of the fluid affects the thickness of the deposited film, and address the discrepancy on the topic in literature. 
\end{abstract}

\maketitle

\section{Introduction}\label{sec_intro}

\begin{figure}[h]
\includegraphics[width = 0.6\textwidth]{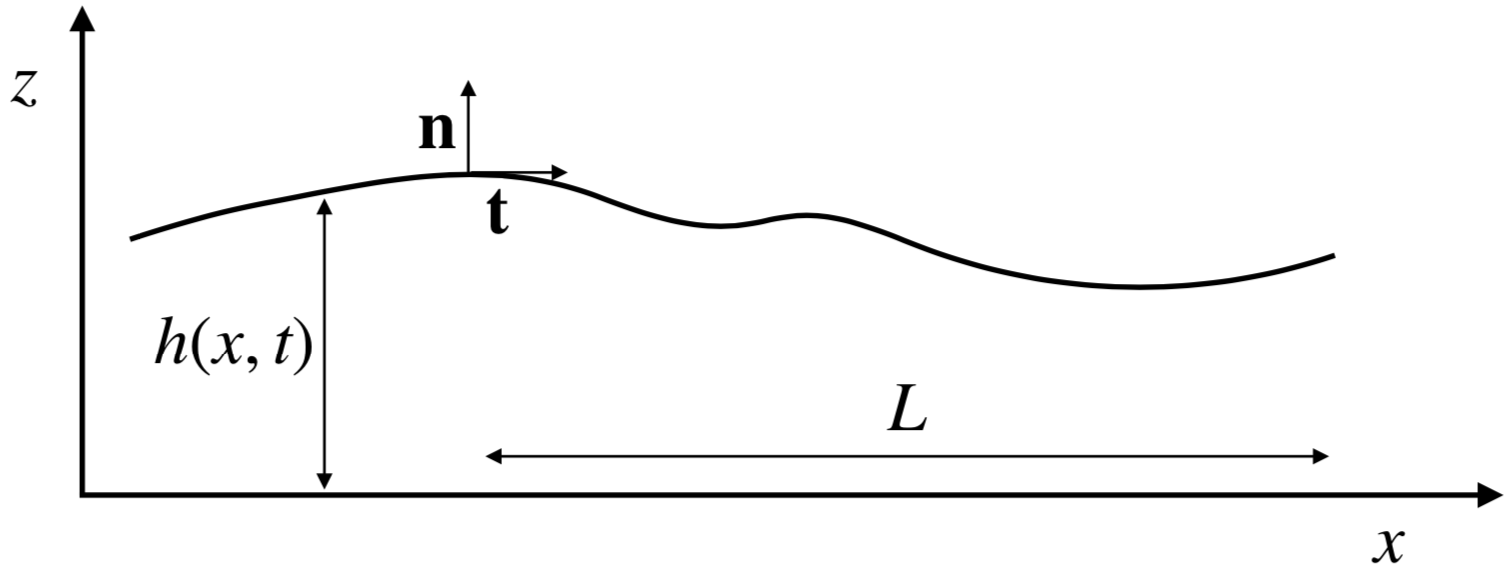}
\caption{Schematic of a thin-film.} 
\label{thinfilmsch}
\end{figure}

Many flows of industrial and biological importance can be characterised as thin-film flows \citep{quere1999, kumar2021actomyosin}. The fluid in a typical free-surface thin-film (see figure \ref{thinfilmsch}) is bounded on one side by a no-slip rigid surface while the opposite surface is free of shear stress; the characteristic length scale in the direction perpendicular to the rigid surface is significantly smaller than along it. Often inertia is also negligible compared to viscous forces, for the magnitude of velocity is often small due to the presence of no-slip surface and the thinness of the film. The disparity in length scales allows for a slender analysis and thereby the Stokes equations are greatly simplified \citep{Oron1997, Eggers2015}. 

The thin-film analysis, or the lubrication approximation, dates back to the work of Reynolds \citep{Reynolds1886} on fluid layers confined between solid boundaries. A similar analysis for films with a free surface \citep{Oron1997} has led to a detailed understanding for a variety of free-surface flows of Newtonian fluids \citep{moffatt1977, Oron1997, homsy2000, craster2009, jacco2013}. One, on which we focus here, is the classical Landau-Levich dip-coating problem \citep{levich1942dragging} where a plate pulled out of a liquid bath is coated by a thin film of the liquid. Many fluids of importance, however, are non-Newtonian and demonstrate properties such as shear-thinning rheology and viscoelasticity. It is therefore of interest to predict thin-film dynamics in fluids with these properties.  In this work, we restrict our attention to viscoelasticity. 

There have been a few previous attempts to understand the effects of viscoelastic fluid properties in flows in thin-films. A thin-film equation for a linear viscoelastic fluid of Jeffreys type was derived in  \citep{rauscher2005thin} where it was used to study the shape of rim in a dewetting film. The model was also used to study viscoelastic effects in  thin-film phenomena such as instability, film rupture, film levelling, and drop spreading \citep{benzaquen2014viscoelastic, tomar2006instability, blossey2012thin, Barra2016}.
The linear Jeffreys model does not have geometric non-linear terms of models like upper-convected Maxwell which ensure frame invariance of the constitutive equation \citep{morozov2015introduction}. This simplicity makes analyses with this model lucrative. The lack of frame-invariance in the model, however, can give rise to unphysical results. For instance, when evaluating a problem in a frame of reference where it is steady \textemdash{} say the Landau-Levich problem in the reference frame of the stationary bath \textemdash{} all viscoelastic terms drop out of the linear constitutive equation. This does not happen in models like the upper-convected Maxwell or the second-order fluid model which are frame-invariant, where the presence and absence of viscoelastic effects is not contingent on the chosen frame of reference.  
 The classical Landau-Levich dip-coating problem \citep{levich1942dragging} as well as the related Bretherton problem \citep{bretherton_1961} were therefore studied for an Oldroyd-B fluid in \citep{ro1995},  using the method of matched asymptotics. The central result of \citep{ro1995} is that weak viscoelasticity reduces the film thickness compared to {that in a Newtonian fluid with the same shear viscosity}.  In contrast, a theoretical and experimental analysis using the Criminale-Ericksen-Filbey (CEF) fluid\textemdash a model fluid valid for weakly viscoelastic flows, found viscoelasticity to increase the film thickness in the coating problem \citep{ashmore2008}. Previous experiments had also found that the film thickness increases compared to the Newtonian case for fluids showing both shear-thinning viscosity and viscoelasticity \citep{de1998fluid}. Numerical simulations using Oldroyd-B and FENE-CR fluids, however, show that weak viscoelasticity decreases film thickness whereas strong viscoelasticity can increase it \citep{Shaqfeh2002}.

In this work, we derive the thin-film equation for a second-order fluid model {\citep{astarita1974principles, morozov2015introduction}}, a common model to study viscoelastic fluids analytically {\citep{leal79, koch_subu, onshun2012}}. Using no approximations besides the slenderness assumption, we perform the long-wave analysis for the second order fluid. This gives a frame-invariant thin-film equation that contains viscoelasticity. We then use the equation to study the Landau-Levich problem and resolve the film thinning vs.\ thickening discrepancy present in the literature.

\section{Theoretical formulation} \label{sec_theory}

We first present the incompressible second-order fluid model that we use to model the viscoelastic fluid, then the long-wave analysis for thin-film flows and finally give the resulting thin-film equations for a second-order fluid.

\subsection{The second-order fluid}

The equations for mass and momentum conservation in the absence of inertia read {\citep{pozrikidis1997introduction}} 
\begin{eqnarray}
\nabla \cdot \mathbf u &=& 0 \\
\nabla \cdot \boldsymbol \sigma &=& \mathbf 0,
\end{eqnarray}
where $\mathbf u$ is the fluid velocity and the stress tensor $\boldsymbol \sigma = - p \mathbf{I} + \boldsymbol{\tau}$, where $p$ is the pressure, $\mathbf{I}$ the identity tensor. The extra stress $\boldsymbol{\tau}$ for a second-order fluid is given by \citep{ astarita1974principles, tanner2000engineering, morozov2015introduction}
\begin{equation}
    \boldsymbol{\tau} = \eta \dot{\boldsymbol{\gamma}} - \dfrac{\Psi_1}{2} \buildrel \nabla \over {\dot{\boldsymbol{\gamma}}} + \Psi_{2} \dot{\boldsymbol{\gamma}}\cdot \dot{\boldsymbol{\gamma}}, 
    \label{eq_S-O}
\end{equation}
where $\eta$ is the fluid viscosity, $\Psi_1$ and $\Psi_2$ are the first and second normal stress coefficients. The upper-convected derivative is defined as {\citep{morozov2015introduction}}
\begin{equation}
    \buildrel \nabla \over {\dot{\boldsymbol{\gamma}}} \equiv \dfrac{\partial \dot{\boldsymbol{\gamma}} }{\partial t} + \mathbf{u} \cdot \nabla \dot{\boldsymbol{\gamma}} - \left( \nabla \mathbf{u} \right)^{T} \cdot \dot{\boldsymbol{\gamma}} - \dot{\boldsymbol{\gamma}} \cdot \nabla \mathbf{u}. 
    \label{uppcon}
\end{equation}
While thermodynamics requires $\Psi_1 < 0$ and $\Psi_1 + 2 \Psi_2 = 0$ when the second-order fluid is considered a fluid in its own right \citep{rajagopal}, these constraints are often not found to be obeyed in flows of many common polymeric fluids \citep{decorato16b}. The second-order fluid, however, can be seen as an approximation (at the second-order) to a more general class of fluids: $\eta, \Psi_1$ and $\Psi_2$ can then uniquely be determined experimentally from viscometric flows of fluids \citep{truesdell}. It is in this latter light that we use the {model; for a recent discussion on the second-order fluid and its parameters, see \citep{christov2016, decorato16b}}. The second-order fluid model has provided insights into many canonical problems in fluid mechanics \citep{koch_subu, leal79, jimmy_second_order}. 

We consider two-dimensional thin films with a free surface at $z=h(x,t)$, flowing over a solid with a no-slip boundary condition at $z=0$. We then have  \citep{Eggers2015}

\begin{eqnarray}\label{eq:slipbc}
\mathbf u&=& \mathbf 0, \; {\text{at}} \; z=0, \\
\boldsymbol \sigma \cdot \mathbf n &=& \gamma \kappa \mathbf n, \; {\text{at}} \;  z=h(x,t). \label{eq:stressbc}
\end{eqnarray}
Here, $\gamma$ is the surface tension and $\kappa$ the interface curvature. Finally, the kinematic equation for the motion {of} the free surface reads {\citep{Eggers2015}},

\begin{equation}
     \dfrac{\partial h}{\partial t}+ \dfrac{\partial }{\partial x} \int_0^h u\: \text{d}z = 0,
    \label{eq:kinematic}
\end{equation}
which is here written in integral form.

\subsection{Long-wave analysis} 

We now perform a long-wave analysis to derive the thin-film equation in two dimensions for an incompressible second-order fluid. To do so, we non-dimensionalise quantities using the familiar Newtonian scalings, namely, heights along the vertical (z-axis) and horizontal (x-axis) directions are scaled by some respective characteristic {scales} $H$ and $L$ (see figure \ref{thinfilmsch}). The flow velocity in the horizontal direction is scaled with $U$, while the vertical velocity is scaled with $\varepsilon U$, where $\varepsilon = H/L$ is assumed to be small due to the thinness of the film \citep{Oron1997, Eggers2015}. Time is scaled by $L/U$, and pressure, using dominant balance, is scaled by $\eta U L/ H^2$. All scaled variables are marked with an overhead bar. We have neglected the role of gravity (or any conservative body force) in the following but it can be included in a straight-forward manner \citep{Oron1997, rauscher2005thin}. 

Next, all flow quantities are expanded as a regular perturbation expansion in $\varepsilon$:
\begin{eqnarray}
\bar u(\bar x,\bar z, \bar t) &=& \bar u_0 + \varepsilon \bar u_1 + \mathcal O(\varepsilon^2) \nonumber \\
\bar w(\bar x,\bar z, \bar t) &=& \bar w_0 + \varepsilon \bar w_1 + \mathcal O(\varepsilon^2)\nonumber \\
\bar p(\bar x, {\bar z}, \bar t) &=& \bar p_0 + \varepsilon \bar p_1+ \mathcal O(\varepsilon^2). 
\label{expansion_eq}
\end{eqnarray}
Substituting \eqref{expansion_eq} in the continuity equation $\nabla \cdot \mathbf u =0$ gives, at leading order, 
\begin{equation}
\frac{\partial \bar u_0}{\partial \bar x} + \frac{\partial \bar w_0}{\partial \bar z} =0,  
\end{equation}
while the momentum conservation $\nabla p =\nabla \cdot \boldsymbol{\tau}$ at leading order becomes:

\begin{eqnarray}
 \frac{\partial \bar p_0}{\partial \bar x} &=&  \frac{\partial^2 \bar u_0 }{\partial \bar z^2} 
+ 
\varepsilon\, Wi \left[2 b \frac{\partial }{\partial \bar x}\left[ \left( \frac{\partial \bar u_0}{\partial \bar z}\right)^2 \right] 
  -  \mathcal L[\bar u_0,\bar w_0] \right]   \label{eq:momx}
 \\
\frac{\partial \bar p_0}{\partial \bar z} &=&  
 2  \varepsilon \, b\, Wi\  \frac{\partial }{\partial \bar z}\left[ \left( \frac{\partial \bar u_0}{\partial \bar z}\right)^2 \right],\label{eq:momy}
\end{eqnarray}
where we have introduced the nonlinear operator

\begin{eqnarray}
    \mathcal{L}\left[\bar u,\bar w \right] 
    &=&  \dfrac{\partial}{\partial \bar t} \left(\dfrac{\partial^2 \bar u}{\partial \bar z^2} \right) 
      + \bar w \left( \dfrac{\partial^3 \bar u}{\partial \bar z^3}\right) 
      + 
     \frac{\partial }{\partial \bar x}\left[  \bar u \frac{\partial^2 \bar u}{\partial \bar z^2}  
     - \frac{1}{2} \left( \frac{\partial \bar u}{\partial \bar z}   \right)^2 \right].
\end{eqnarray}
The equations contain two dimensionless parameters, the Weissenberg number
\begin{equation}
Wi = \frac{U \Psi_{1} }{2 \eta H}, 
\end{equation}
and the ratio of the two normal-stress-difference coefficients $b = \Psi_2/ \Psi_1 $. The factor of $1/2$ in the Weissenberg number defined above ensures equivalence with its definition in other common models such as upper-convected Maxwell and Giesekus where instead of the first-normal stress coefficient the relaxation time $\lambda$ of the fluid is used \citep{morozov2015introduction, onshun2012, decorato16b}. We note that the terms that effect viscoelasticity in the problem are retained at the leading order under the assumption that $\mathcal{O}\left(\varepsilon Wi \right) \sim \mathcal{O} \left( 1\right) \gg \mathcal{O}\left( \varepsilon^2\right)$.

We now turn to the boundary conditions. At the solid {wall, we} simply have $\bar u_0=\bar w_0=0$. {At} leading order, the normal component of the stress boundary condition (\ref{eq:stressbc}) at the free-surface $\bar z = \bar h(\bar x,\bar t)$ takes the form 
\begin{equation}
       - \bar p_0 + { 2 \varepsilon\ b\ Wi \left(\dfrac{\partial \bar u_0}{\partial \bar z} \right)^2} = \frac{\epsilon^3}{Ca} \, \dfrac{\partial^2 \bar h}{\partial \bar x^2},
    \label{eq:normal_stress}
\end{equation}
where we have defined the capillary number 
\begin{equation}
Ca = \frac{\eta U}{\gamma}. 
\end{equation}
As in the standard Newtonian lubrication theory, this boundary condition reveals that $\varepsilon = \mathcal O\left(Ca^{1/3}\right)$ for a leading-order dependence of pressure on the curvature of the film, and therefore the expansion parameter $\varepsilon$ may be chosen equal to $Ca^{1/3}$ disregarding the proportionality constant {\citep{Eggers2015}}. The tangential component of the stress boundary condition (\ref{eq:stressbc}) at the free-surface $z = h(x,t)$, {at} leading order, is 
\begin{equation}
    \eta \dfrac{\partial \bar u_0}{\partial \bar z} - {\varepsilon \, Wi\, \mathcal{M}\left[\bar u_0,\bar w_0\right]} = 0,
    \label{eq:tangential_stress}
\end{equation}
where 
\begin{equation}
    \mathcal{M}\left[\bar u,\bar w \right] =  \dfrac{\partial}{\partial \bar t} \dfrac{\partial \bar u}{\partial \bar z}+  \bar u \dfrac{\partial^2 \bar u}{\partial \bar x\:\partial \bar z} + \bar w  \dfrac{\partial^2 \bar u}{\partial \bar z^2} + 2 \dfrac{\partial \bar u}{\partial \bar z}\dfrac{\partial \bar u}{\partial \bar x} + 2 \dfrac{\partial \bar h}{\partial \bar x} \left(\dfrac{\partial \bar u}{\partial \bar z}\right)^2.
    \label{formathcalM}
\end{equation}
We remark that all equations and boundary conditions only involve the leading order quantities $\bar u_0,\bar w_0,\bar p_0$.

\subsection{Viscoelastic thin-film equation}

We briefly recall the usual Newtonian lubrication limit (i.e. $Wi=0$). The vertical momentum \eqref{eq:momy} famously gives $\partial \bar p/\partial \bar z=0$, so that {the} pressure is homogeneous across the thickness of the flowing layer. This combined with the horizontal momentum equation \eqref{eq:momx} gives the velocity its parabolic (Poiseuille) structure, 
\begin{equation}\label{eq:velocity}
\bar u_0(\bar x, {\bar z},\bar t) =-\dfrac{\bar A\left(\bar x, \bar t\right)}{2} \left( \bar z^2 - 2\bar h \bar z\right),
\end{equation}
for which the no-slip (at $z=0$) and no-shear {(at $z=h$)} boundary conditions  have been used. The kinematic condition of the interface (\ref{eq:kinematic}) then becomes
\begin{equation}\label{eq:flux}
\frac{\partial \bar h}{\partial \bar t} + \dfrac{1}{3} \frac{\partial }{\partial \bar x}\left( \bar A \bar h^3\right) = 0.
\end{equation}
The function $\bar A(\bar x, \bar t)$ reflects the strength of the flow {and in} the Newtonian {case
is} simply proportional to the pressure gradient.

Our aim is to study the viscoelastic effects originating from $\Psi_{1,2}$ in \eqref{eq_S-O}. The momentum equations (\ref{eq:momx},\ref{eq:momy}) are now more intricate compared to the Newtonian case, but can still be solved upon recognising that the viscoelastic terms are nearly in the form of a gradient.  Indeed, a redefinition of pressure
\begin{equation}
\bar p_{\rm mod} = \bar p_0 + { \varepsilon \,Wi} \left(\bar u_0 \dfrac{\partial^2 \bar u_0}{\partial \bar z^2}-\dfrac{1}{2}\left(\dfrac{\partial \bar u_0}{\partial \bar z}\right)^2  \right)  - 2 b\, \varepsilon\, Wi \left(\dfrac{\partial \bar u_0}{\partial \bar z}\right)^2 ,
\end{equation}
turns the momentum equations \eqref{eq:momx} and \eqref{eq:momy} into
\begin{eqnarray}
 \frac{\partial \bar p_{\rm mod}}{\partial \bar x} &=&  
\left( 1 -  \varepsilon\, Wi\, \frac{\partial }{\partial \bar t}\right) \left( \frac{\partial^2 \bar u_0}{\partial \bar z^2} \right) 
-   \varepsilon\, Wi\, \bar w_0 \frac{\partial^3 \bar u_0}{\partial \bar z^3},  \label{eq:mombisx} \\
\frac{\partial \bar p_{\rm mod}}{\partial \bar z} &=&  
   \varepsilon\, Wi\, \bar u_0 \frac{\partial^3 \bar u_0}{\partial \bar z^3}. \label{eq:mombisy} 
\end{eqnarray}
The key observation is that the parabolic form of the velocity profile given in \eqref{eq:velocity} is a solution to the set of equations for the viscoelastic fluid. To see this, we first note that the velocity  \eqref{eq:velocity} still satisfies the tangential shear-stress condition at the free-surface \eqref{eq:tangential_stress} (once the kinematic condition at the free-surface is taken into account).  
Then we note that the vertical momentum equation \eqref{eq:mombisy} for the parabolic velocity gives the modified pressure a constant value across the film thickness, i.e. $\partial \bar p_{\rm mod}/\partial \bar z=0$. The pressure gradient, consistently, follows from the horizontal momentum equation (\ref{eq:mombisx})
\begin{equation}\label{eq:momtresx}
 \frac{\partial \bar p_{\rm mod}}{\partial \bar x} = -\left( 1 - \varepsilon\, Wi\, \frac{\partial }{\partial \bar t} \right) \bar A\left(\bar x, \bar t \right),
\end{equation}
as independent of $z$. Finally, the normal stress boundary condition \eqref{eq:normal_stress} gives another expression for the pressure
\begin{equation}\label{eq:bcpmod}
\bar p_{\rm mod} =  -\frac{\varepsilon\, Wi }{2}\bar A^2 \bar h^2   - \frac{\epsilon^3}{Ca} \frac{\partial^2 \bar h}{\partial \bar x^2}.
\end{equation}
Together with \eqref{eq:momtresx}, this gives the sought-after relation, with viscoelastic effects, between the flow strength $\bar A$ and the gradient of capillary pressure that drives the {flow:}
\begin{equation}
 \frac{\epsilon^3}{Ca} \frac{\partial^3 \bar h}{\partial \bar x^3} - 
 \left( 1 - {\varepsilon\, Wi} \frac{\partial }{\partial \bar t} \right) \bar A + \frac{\varepsilon\, Wi}{2} \frac{\partial }{\partial \bar x}\left(\bar A^2 \bar h^2\right) =0.
 \label{mainresult}
\end{equation}
Equations \eqref{mainresult} and \eqref{eq:flux} form a closed set of equations for the height $\bar h$ of the film and the flow strength $\bar A$ in the film. Note that of the two normal stress difference coefficients, it is only the first that enters the equations through $Wi$. 

In the aforementioned, we have seen the parabolic velocity profile satisfy the equations for a second-order fluid.  It has not escaped our attention that this is reminiscent of Tanner's theorem \citep{tanner_theorem, huilgol1973uniqueness} which states that a two-dimensional flow of a second-order fluid admits a Newtonian velocity field with given velocity boundary conditions. Importantly, however, free surface flows involve \emph{stress} boundary conditions at the free-surface and hence the applicability of Tanner's theorem is uncertain a priori. Indeed, the velocity field (\ref{eq:velocity}) for the second order fluid is not identical to the Newtonian case, for $\bar A(\bar x, \bar t)$ in the two fluids will be different although the $z$-dependence is preserved.

\subsection{Summary}

\subsubsection{Dimensional equations} 

We now summarise the result of the long-wave expansion and discuss various forms of the thin film equations. To facilitate a physical interpretation, the results will be presented in dimensional form. A key observation is that, like in the Newtonian lubrication approximation, the leading order velocity in the second order fluid is still parabolic:
\begin{equation}\label{eq:velocitybis}
u(x,z,t) =-\dfrac{A\left(x,t\right)}{2} \left(  z^2 - 2 z h(x,t)  \right).
\end{equation}
The thin film description then consists of two coupled equations for the interface shape $h(x,t)$ and the flow strength $A(x,t)$:
\begin{eqnarray}
\dot{h} + \dfrac{1}{3} \left( A h^3\right)' &=& 0, \label{eq:rig_so1} \\
\gamma h''' -  \eta A + \dfrac{\Psi_1}{2} \left[ \dot{A} + \dfrac{1}{2} \left( A^2 h^2\right)'\right] &=&0,
\label{eq:rig_so2}
\end{eqnarray}

where dots and primes denote derivatives with respect to $t$ and $x$. The former of these equations represents mass conservation. The latter equation represents a momentum balance that relates the flow strength $A$ to driving by surface tension ($\gamma$) in the presence of viscosity ($\eta$) and first normal stress difference {coefficient} $(\Psi_1)$. Interestingly, the second normal stress difference {coefficient} completely drops out of the expansion {at leading order and} does not appear in the thin film equations.

The thin-film equations above can easily be extended to include the presence of a conservative body force with potential $\phi\left(x, z\right)$ as well as when the Navier slip boundary condition apply at the solid wall i.e. at $z=0$, $u = \ell_s \partial u/\partial z$, where $\ell_s$ is the slip-length {\citep{Oron1997}}. With the extension, they become 
\begin{eqnarray}
\dot{h} + \dfrac{1}{3} \left( A h^2 \left(h + 3\ell_s \right)\right)' &=& 0, \label{eq:rig_so_main1}  \\
\gamma h''' - \phi' \left(x, h\left(x, t \right)\right) - \eta A + \dfrac{\Psi_1}{2} \left[ \dot{A} + \dfrac{1}{2} \left( A^2 h \left(h + 2\ell_s \right)\right)'\right] &=& 0.
\label{eq:rig_so_main2}
\end{eqnarray}
Note that in the absence of viscoelasticity, i.e. without $\Psi_1$ in the equations, one recovers the thin-film equation in Newtonian fluids by substituting the expression of $A$ from \eqref{eq:rig_so_main2} in \eqref{eq:rig_so_main1} {\citep{Oron1997}}.

\subsubsection{Steady flows} 

The thin-film equations simplify when one is looking for traveling wave solutions for which profiles propagate steadily with some constant velocity, say $U$, with respect to the solid boundary.  It is worth re-emphasising here that a second-order fluid model, seen as a slow-flow approximation, is inappropriate for strongly unsteady problems \citep{astarita1974principles, morozov2015introduction}. From this perspective, the steady travelling wave solutions perhaps form the most important applications of our study. When imposing the form $h(x-Ut)$ and $A(x-Ut)$ in (\ref{eq:rig_so1}), the flow strength $A$ now expressed in a comoving coordinate $\tilde{x} \equiv x - U t$ is
\begin{equation}
A = \frac{3U(h-h^*)}{h^3},
\end{equation}
where $h^*$ is an integration constant that controls the flux in the flowing layer \citep{Eggers2015}. Substituting this for $A$ in (\ref{eq:rig_so2}), the ordinary differential equation for the profile in the co-moving frame becomes
\begin{equation}
\gamma h''' - \dfrac{3\eta U}{h^3} \left( h - h^*\right) - \dfrac{3 \Psi_1 U^2 h' }{2h^3} F \left(\frac{h}{h^*}\right) = 0,
\label{eq:LL}
\end{equation}
with
\begin{equation}\label{eq:Fcharu}
F(x) =  1 - 6 x^{-1} + 6x^{-2}.
\end{equation}
Equation \eqref{eq:LL} will be discussed in detail below but it is immediately clear that for $\Psi_1 = 0$ one obtains the equation used for the classical Landau-Levich problem \citep{levich1942dragging}. In section~\ref{sec:LL} we address how viscoelasticity affects the thickness of a liquid film that is entrained via a dip-coating experiment, and we will compare our results to previous approaches in the literature. A special case of (\ref{eq:LL}) case is when $h^*=0$, for which the depth-integrated flux across the layer vanishes in the co-moving frame. This form of the thin-film equation has been extensively studied in the context of moving contact lines \citep{Eggers2015}. The equation for a moving contact line with viscoelasticity thus reads
\begin{equation}
\gamma h''' - \dfrac{3\eta U}{h^2} - \dfrac{3 \Psi_1 U^2 h' }{2h^3} = 0.
\label{eq:contactline}
\end{equation}
The equation is similar to the viscoelastic contact line equation proposed in \citep{boudaoud2007} based on scaling estimates for components of the stress tensor. While \citep{boudaoud2007} correctly captures the scaling $\Psi_1 U^2h'/h^3$ of the viscoelastic term its estimation of the prefactor to be $18$ is different from the value $3/2$ obtained through the long-wave expansion here.\\

\section{The viscoelastic Landau-Levich problem}\label{sec:LL}

\begin{figure}[h]
\includegraphics[width = 0.4\textwidth]{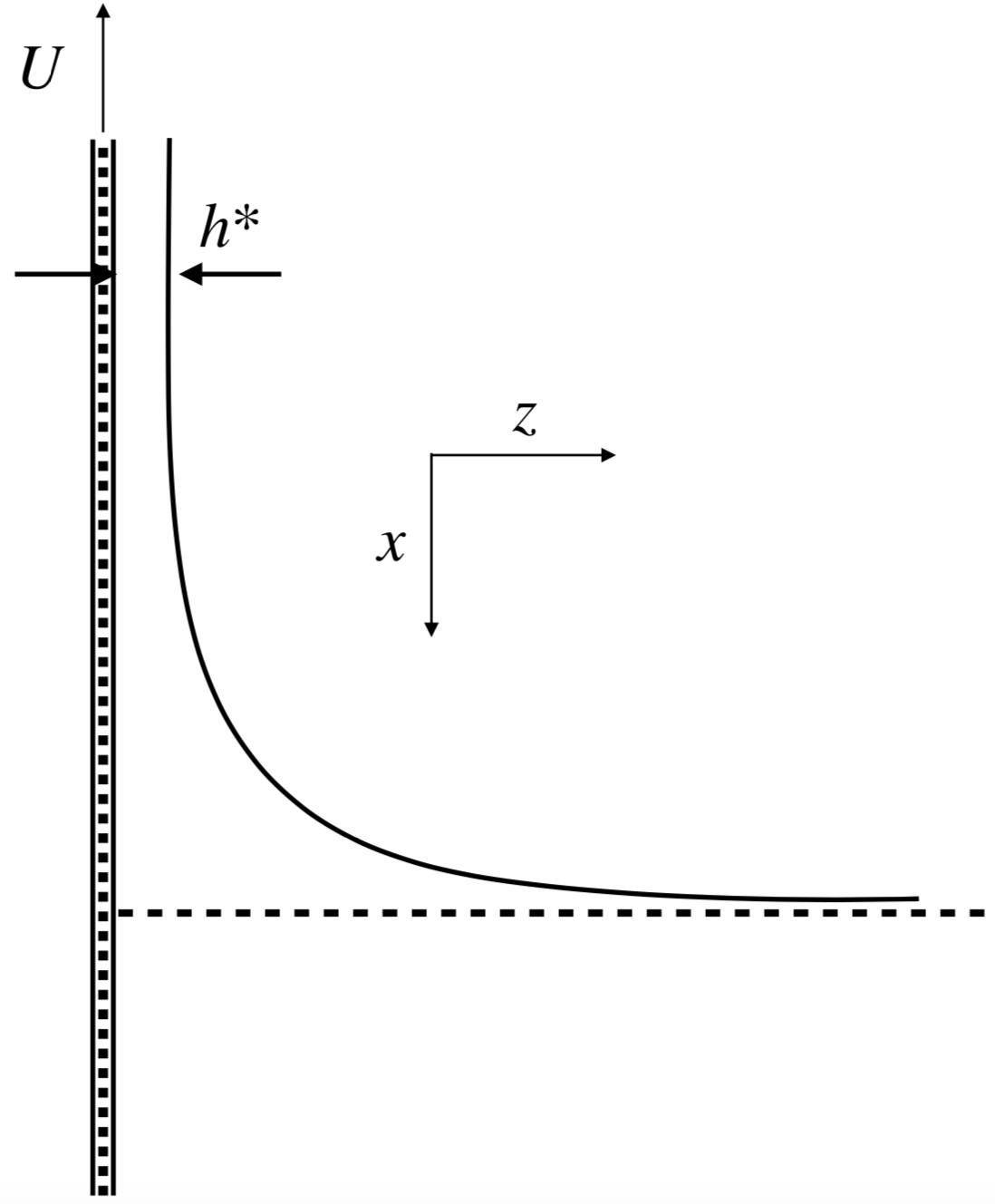}
\caption{Schematic of the Landau-Levich dip coating problem.} 
\label{LLsch}
\end{figure}

We now use the thin-film equation derived in the aforementioned to compute the effect of viscoelasticity on the Landau-Levich problem where a film is entrained by pulling a solid plate from a quiescent liquid bath as shown in the schematic of figure~\ref{LLsch}. We assume the film to be steady in the lab frame (frame of bath), which in the frame of reference attached to the plate corresponds to travelling wave solutions. These are described by \eqref{eq:LL}, where $U$ now is the imposed plate velocity. The negative $x$ direction is up along the plate and as $x \rightarrow -\infty$ a homogeneous thickness $h \rightarrow h^*$ coats {the plate}. The coated film merges into the stationary bath where $h \rightarrow \infty$. This latter boundary condition is established through the technique of matched asymptotics by matching the film curvature to that of the stationary bath \cite{wilson1982drag}
\begin{equation}\label{eq:curvature}
h'' \to \dfrac{\sqrt{2}}{\ell} \quad \text{as} \quad h \to \infty, 
\end{equation}
where {$\ell = \sqrt{\gamma/ \rho g}$ is} the capillary {length, $\rho$} being the density of the {fluid and} $g$ the acceleration due to gravity \citep{wilson1982drag}. The Landau-Levich scaling for the entrained thickness in Newtonian fluids is $h^* \sim \ell Ca^{2/3}$ {at} leading order \citep{wilson1982drag, levich1942dragging}. The scaling also appears in other entrainment problems \citep{stone_2010} like pulling of a soap film \citep{nierop2008, quere1999} or motion of a long bubble in a tube \textemdash the Bretherton problem \cite{bretherton_1961}. The ensuing analysis in these problems is similar and details particular to the problem often present themselves mainly through the length scale $\ell$ appearing in the scaling relation, or at higher orders \citep{park_homsy_1984}. Here the goal is to determine how $h^*$ relates to the plate velocity in the presence of viscoelasticity. Note that in this work we do not consider the thin-film dynamics near the wetting transition, where film thicknesses other than the Landau-Levich thickness have been known to exist in Newtonian fluids \citep{Jacco2008, uwe2014, jimmy2016}.

A natural dimensionless parameter to quantify viscoelastic effects would be $\Psi_1 U/(2\eta h^*)$ \textemdash a local Weissenberg number, comparing the timescale $\Psi_1/2\eta$ to the local shear-rate $U/h^*$. However, $h^*$ is not known a priori and we therefore define the Weissenberg number 
\begin{equation}\label{eq:Wi}
Wi_{\ell} = \frac{\psi_1 U}{2\eta \ell},
\end{equation}
based on the capillary length $\ell$ which appears in the curvature boundary condition (\ref{eq:curvature}).

We now proceed by introducing scales from the Newtonian Landau-Levich problem so that
\begin{equation}
h(x) = \ell Ca^{2/3} H(X), \quad X = \frac{x}{\ell Ca^{1/3}},
\end{equation}
which transform (\ref{eq:LL}) to 
\begin{equation}\label{eq:LLscaled}
 H''' - \dfrac{3(H-H^*)}{H^3} - 3 Wi_{\ell} \, Ca^{-1/3} \dfrac{H'}{H^3} F(H/H^*)= 0.
\end{equation}
The film-thickness $H^*$ that is to be determined is selected by the unique solution that satisfies the boundary conditions at the two ends of the plate, namely, at the region of a flat film, and where the film meets the bath: 
\begin{equation}\label{eq:LLbc}
H(-\infty) \to H^*, \quad \quad H''(\infty) \to \sqrt{2}. 
\end{equation}
For the Newtonian case it is known that {$H^*=0.946$} \citep{wilson1982drag, levich1942dragging}, but now with viscoelasticity the entrained film thickness will be function of the parameter $Wi_{\ell} \, Ca^{-1/3}$. 

\begin{figure}
\includegraphics[width = 0.66\textwidth]{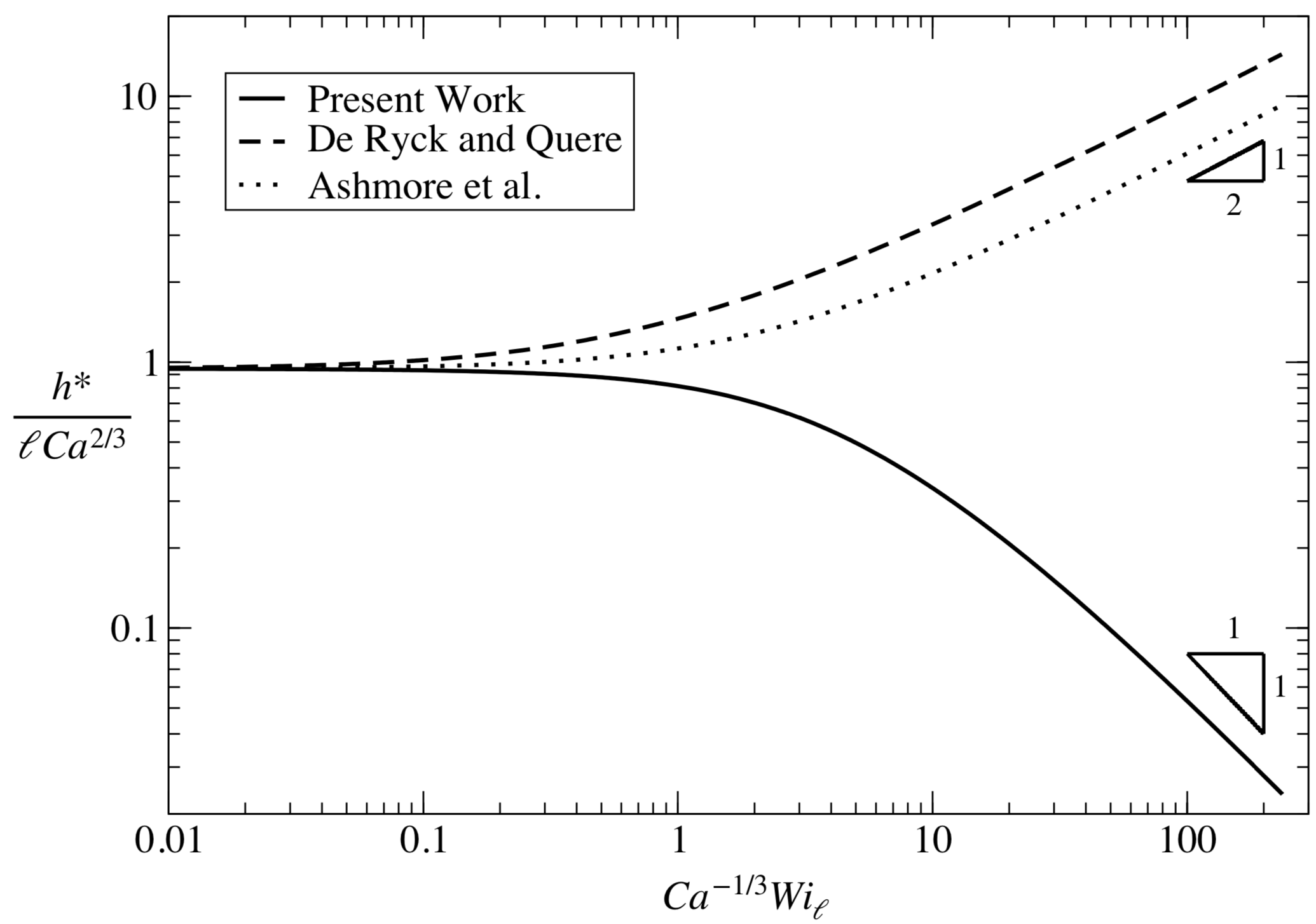}
\caption{Thickness of the Landau-Levich film $h^*$, {compensated by the Newtonian scaling}, in the presence of viscoelasticity. We plot the results by solving numerically our equation \eqref{eq:LLscaled} and those from other thin-film equations \citep{de1998fluid, ashmore2008}. See text for a detailed discussion of the literature. 
}
\label{LL_figall}
\end{figure}

The boundary value problem defined by (\ref{eq:LLscaled},\ref{eq:LLbc}) is solved numerically, and the result is shown as the solid line in figure~\ref{LL_figall}. At small values of $Wi_{\ell} \, Ca^{-1/3}$, one recovers the Newtonian value {$H^*=0.946$}. The effect of viscoelasticity in the second order fluid is to decrease the film thickness with respect to the Newtonian value. In fact at values of $Wi_{\ell} \, Ca^{-1/3} \ll 1$, we find [details in Appendix]
\begin{equation}\label{eq:small}
\dfrac{h^*}{\ell Ca^{2/3}} = 0.946  - 0.138 \, Wi_{\ell} \, Ca^{-1/3} + \ldots \, .
\end{equation}
Even in the limit where $Wi_{\ell} \, Ca^{-1/3} \gg 1$, the thickness continues to decrease (figure~\ref{LL_figall}). The numerical solution suggests a scaling $H^* \sim  (Wi_{\ell} \, Ca^{-1/3})^{-1}$. This decrease at large $Wi_{\ell} \, Ca^{-1/3}$ contradicts $H^* \sim  (Wi_{\ell} \, Ca^{-1/3})^{1/2}$ \textemdash{} in dimensional terms $h \sim U$ \textemdash{} found in previous Landau-Levich analyses that include viscoelasticity  \citep{de1998fluid,ashmore2008} (see comparison in figure~\ref{LL_figall} and details in {Appendix}). The reason for this difference and a summary of results in the literature is discussed next.

\subsection{Note on previous literature}

The seminal analytical work on the effect of viscoelasticity on the Landau-Levich film thickness came from \citet{ro1995} who studied the problem using matched asymptotics in an Oldroyd-B fluid. In brief, they performed a double expansion of flow quantities in $Ca^{1/3}$ and $Wi_{\ell}\, Ca^{-1/3}$, and found that when these quantities are small the film thickness decreases compared to the Newtonian value. Their result is similar to our \eqref{eq:small} but differs in the numerical coefficient of correction ($Wi_{\ell}\, Ca^{-1/3}$) term. We suspect the difference to be a typo or a numerical error, since an expansion of (\ref{eq:LL}) in the small $Wi_{\ell}\, Ca^{-1/3}$ limit gives the exact same equations as in \citep{ro1995} {(see Appendix for details).} This equivalence at low $Wi$ is a direct consequence of the second-order fluid being a low $Wi$ limit of upper-convected {Maxwell} (with $\Psi_2 = 0$) and Giesekus models; a recent demonstration of this was shown by \citet{decorato16b}. 

A decrease in film-thickness with viscoelasticity was also observed for the closely related Bretherton problem \citep{bretherton_1961} (the Landau-Levich and Bretherton problems scale similarly with $Ca$ at the leading order in Newtonian fluids \citep{park_homsy_1984, wilson1982drag}) by the numerical simulations of \citet{Shaqfeh2002} for Oldroyd-B and FENE-CR fluids. They found that the film thickness at low $Wi$ and $Ca$ initially decreases compared to the Newtonian values, consistent with \citep{ro1995} and the present study. However, the film thickness in the numerical simulations of \citet{Shaqfeh2002} was found to increase compared to the Newtonian case at larger values of $Wi$ and $Ca$. Clearly, this increase is not captured by our thin-film equation using the second order fluid (see figure~\ref{LL_figall}). We remind the reader that the second-order fluid model is not an appropriate model at relatively large $Wi$ because it is a model formally derived for slow flows \citep{astarita1974principles}. It is not surprising, therefore, that our results do not capture this increase in film thickness. 

The increase in film thickness was, however, found in the experiments and analysis of \citet{de1998fluid} for a shear-thinning viscoelastic fluid. The work devised a thin-film equation not from a specific constitutive relation but prompted by estimates for the components of the stress tensor \citep{de1998fluid}. The proposed lubrication equation (for the case without shear-thinning) has the same structures as (\ref{eq:LL}), but, importantly, the function $F$ is different: 
\begin{equation}\label{eq:quere}
F (x) =  4\frac{(x-1)(2x-3)}{x^2}.
\end{equation}
This change in $F$, compared to our \eqref{eq:Fcharu}, has a significant consequence for the Landau-Levich film thickness, as can be seen in figure \ref{LL_figall}. While the model of \citep{de1998fluid} captures the increase in film thickness observed experimentally and in numerical simulations, it misses the decrease in film thickness for weak viscoelasticity found in \citep{ro1995, Shaqfeh2002} and the present work.

More recently, \citet{ashmore2008} studied the coating problem for a Criminale-Ericksen-Filbey (CEF) fluid. This fluid model, without shear-thinning i.e. setting power-law index $n=1$, is strictly identical to the second-order fluid. The thin-film equation (for $n=1$) in  \citep{ashmore2008} again has the same structure as (\ref{eq:LL}), but with 
\begin{equation}\label{eq:ashmore}
F(x) = \frac{(3x-4)(x-2)}{x^2}.
\end{equation}
Similar to the work by \citet{de1998fluid}, this equation also predicts an increase of the Landau-Levich film thickness (see figure \ref{LL_figall}). 
We remark, however, that the equivalence between the CEF fluid at $n=1$ and the second-order fluid warrants that the thin-film equation from \citep{ashmore2008} {be} identical to that of the present work. It turns out that in \citep{ashmore2008}, the closure of the analytical problem involved a depth-averaging under an approximation. However, the depth-averaging can be carried out without any approximation, and in that case one recovers (\ref{eq:Fcharu}) rather than (\ref{eq:ashmore}). 

We thus conclude that the lubrication theory for the second-order fluid/CEF fluid at $n=1$ gives a Landau-Levich film thickness that is always thinner than {in a Newtonian fluid with same shear viscosity}. This {establishes} the thinning of the Landau-Levich film in the weakly viscoelastic limit, which is the range where the the second-order fluid is expected to be valid. However, the increase in the film thickness at larger values of $Wi$ observed experimentally and numerically remains to be explained analytically.

\section{Conclusion} \label{sec_conclu}

We have obtained a closed set of thin-film equations \eqref{eq:rig_so_main1} and \eqref{eq:rig_so_main2} for a viscoelastic fluid using the second-order fluid model. This work is a major step forward in obtaining the equation with normal stress differences and offers a gateway to understand their role in free-surface problems. Previous {studies that proposed} similar equations were based on a linear viscoelastic fluid \citep{rauscher2005thin}, which is not frame-invariant and lacks normal-stress feature of viscoelastic fluids; used scaling arguments to derive them \citep{de1998fluid, boudaoud2007} or included additional assumptions including steadiness of flow \citep{ro1995, ashmore2008}. The derived equations \eqref{eq:rig_so_main1} and \eqref{eq:rig_so_main2} may be used to study various free-surface flow problems including contact line dynamics of viscoelastic fluids -- phenomena such as spreading of drops and dewetting. Here, we have used the equations to study the classical Landau-Levich problem, and shown that, in fact, weak viscoelasticity decreases the thickness of the deposited film compared to its Newtonian value. With this result, we have addressed the discrepancy in literature on the effect of  viscoelasticity on the film thickness in the Landau-Levich problem.  

We note that the second-order fluid model studied as a slow-flow approximation should be used only for weakly viscoelastic fluids. The numerical simulations of \citet{Shaqfeh2002} show that at higher values of Weissenberg and {capillary numbers viscoelasticity, in fact, increases} the film thickness compared to {a Newtonian fluid with same shear viscosity}.  It would be interesting to have a thin-film equation corresponding to the strongly elastic limit. We aim to work towards it in the future. 

{{\bf Acknowledgments.} The authors thank {Jens} Eggers for discussions. We acknowledge
support from NWO through VICI grant no. 680-47-632.}

\appendix 

\section{Asymptotics}

\subsection{The Landau-Levich film thickness when $Ca^{-1/3} Wi_{\ell} \ll 1$.}

The thin-film equation for the Landau-Levich problem in scaled variables \eqref{eq:LLscaled} is 

\begin{equation}\label{eq:LLscaledAPP}
 H''' - \dfrac{3(H-H^*)}{H^3} - 3 Wi_{\ell} \, Ca^{-1/3} \dfrac{H'}{H^3} F\left(\dfrac{H}{H^*}\right)= 0,
\end{equation}
with the following boundary conditions
\begin{equation}\label{eq:LLbcAPP}
H(-\infty) \to H^*, \quad \quad H''(\infty) \to \sqrt{2}. 
\end{equation}
Here 
\begin{equation}\label{eq:FcharuAPP}
F(x) =  1 - 6 x^{-1} + 6x^{-2}.
\end{equation}
When $\delta = Ca^{-1/3} Wi_{\ell} \ll 1 $, one may write $H$ and $H^*$ as a regular perturbation expansion in the small variable $\delta$ i.e. 
\begin{equation}
\left\{H,\, H^* \right\}= \left\{H_0,\, H^*_0 \right\} + \delta\, \left\{H_1,\, H^*_1 \right\} + \delta^2\, \left\{H_2,\, H^*_2 \right\} \ldots
\end{equation}
With this, at $\mathcal{O}\left( 1\right)$, the equation and boundary conditions become
\begin{equation}
 H'''_0 - \dfrac{3(H_0-H^*_0)}{H_0^3} = 0,
\end{equation}
\begin{equation}
H_0(-\infty) \to H_0^*, \quad \quad H''_0(\infty) \to \sqrt{2}, 
\end{equation}
which correspond to the Newtonian Landau-Levich problem \citep{levich1942dragging}. At $\mathcal{O}\left(\delta \right)$, we have 
\begin{equation}\label{small_delta1}
 H'''_1 + \dfrac{3 H^*_1}{H_0^4} +  \dfrac{3 H^*_1(2H_0- 3H_0^*)}{H_0^4} - 3 \dfrac{H_0'}{H_0^3} F\left( \dfrac{H_0}{H_0^*}\right)  = 0,
\end{equation}
and 
\begin{equation} \label{small_delta2}
H_1(-\infty) \to H_1^*, \quad \quad H''_1(\infty) \to 0. 
\end{equation}
Equation \eqref{small_delta1} written above and equation 85 in \citep{ro1995} are identical and admit same boundary {conditions} \eqref{small_delta2}.   On solving these equations we find
\begin{equation}
H^*_0 = 0.946; \quad H^*_1 = 0.138,
\end{equation}
which gives the result \eqref{eq:small} presented in the main text. 

\subsection{The Landau-Levich film thickness when $Ca^{-1/3} Wi_{\ell} \gg 1$ }

When $\delta = Ca^{-1/3} Wi_{\ell} \gg 1$ in \eqref{eq:LLscaled} is large, the dominant terms of the equation are expected to be the capillary pressure gradient term and the viscoelastic term. This means $\eqref{eq:LLscaled}$ at leading order is
\begin{equation} \label{elastic_dom}
 H'''  - 3 \delta\, \dfrac{H'}{H^3} F\left(\dfrac{H}{H^*}\right)= 0,
\end{equation}
where the viscous term has been dropped. 
Equation \eqref{elastic_dom} can be readily integrated once to give
\begin{equation}
H''  + \dfrac{3\delta}{H^2} G\left( \dfrac{H}{H^*}\right) + K = 0
\end{equation}
where $K$ is the integration constant, and for $F$ given by {\eqref{eq:LLbcAPP}}
\begin{equation}\label{G-us}
G\left(x\right) = \dfrac{1}{2} - 2 x^{-1} + \dfrac{3}{2} x^{-2}.
\end{equation}
The different expressions of $F\left( x\right)$ in \eqref{eq:quere} and \eqref{eq:ashmore} from the work of \citep{de1998fluid} and \citep{ashmore2008}  result in different $G\left(x\right)$.
To find the constant $K$, we note that as $H \to H^*$, $H'' \to 0$, and this gives $K = -\dfrac{3\delta}{H^{*2}} G\left(1\right)$. Towards the stationary bath, as $H \to \infty$, we find
\begin{equation}\label{strongVEscale}
H''\left( \infty\right) = - K = \sqrt{2},
\end{equation}
where in the first equality we have used the fact that $G\left(\infty\right)$ remains finite, and the second equality uses \eqref{eq:LLbc}. Using the value of $K$ in the second equality of \eqref{strongVEscale} we get
\begin{equation}\label{scaling_SVE}
H^* =\left( \dfrac{3 \delta}{\sqrt{2}} G \left( 1\right)\right)^{1/2}.
\end{equation}
The scaling in this relation is identical to that obtained in \citep{ashmore2008} and \citep{de1998fluid}. Importantly, however, for our expression of $F$, $G\left(1 \right) = 0$ (using \eqref{G-us}) so that \eqref{scaling_SVE} predicts a zero film-thickness. Note that $G\left(1 \right) = 0$ results in $K = 0$ which implies that the curvature boundary condition in \eqref{strongVEscale} cannot be satisfied by just considering the viscoelastic and the capillary term; the viscous term dropped in \eqref{elastic_dom} is therefore important here and must be included to prevent the breakdown of \eqref{scaling_SVE}. 

In contrast, the approximate expressions for $F$ in the works of \citep{ashmore2008} and \citep{de1998fluid}  give a non-zero $G\left(1\right)$. This explains the differences in scaling laws for large $\delta \equiv Ca^{-1/3} Wi_{\ell}$ as observed in figure 3.

\bibliographystyle{jfm}
\bibliography{Lubri_SOF}

\end{document}